Version 4

# Consciousness, cognition, and context: extending the global neuronal workspace model

Rodrick Wallace
The New York State Psychiatric Institute
*

January 21, 2004

## Abstract

We adapt an information theory analysis of interacting cognitive biological and social modules to the problem of the global neuronal workspace, the current standard neuroscience picture of consciousness. Tunable punctuation emerges in a natural manner, suggesting the possibility of fitting appropriate phase transition power law, and, away from transition, generalized Onsager relation expressions, to observational data on conscious reaction. The development can be extended in a straightforward manner to include the role of psychosocial stress, culture, or other embedding structured contexts in individual consciousness, producing a 'biopsychosocial' model that closely retains the flavor of the standard treatment, but better meets compelling philosophical and other objections to brain-only descriptions.

**Key words**: asymptotic limit theorems, cognition, consciousness, Dretske, information theory, Onsager relations, phase transition, renormalization.

## Introduction

A recent special issue of *Cognition* (**79**(1-2), 2001)) explores contemporary work on consciousness in humans, presenting various aspects of the new 'standard model' synthesized over the last decade or so (esp. Dehaene and Naccache, 2001). Sergeant and Dehaene (2004) describe that work, and some of the implicit controversy, as follows:

> "[A growing body of empirical study shows] large all-or-none changes in neural activity when a stimulus fails to be [consciously] reported as compared to when it is reported... [A] qualitative difference between unconscious and conscious processing is generally expected by theories that view recurrent interactions between distant brain areas as a necessary condition for conscious perception... One of these theories has proposed that consciousness is associated with the interconnection of multiple areas processing a stimulus by a [dynamic] 'neuronal workspace' within which recurrent connections allow long-distance communication and auto-amplification of the activation. Neuronal network simulations... suggest the existence of a fluctuating dynamic threshold. If the primary activation evoked by a stimulus exceeds this threshold, reverberation takes place and stimulus information gains access, through the workspace, to a broad range of [other brain] areas allowing, among other processes, verbal report, voluntary manipulation, voluntary action and long-term memorization. Below this threshold, however, stimulus information remains unavailable to these processes. Thus the global neuronal workspace theory predicts an all-or-nothing transition between conscious and unconscious perception... More generally, many non-linear dynamical systems with self-amplification are characterized by the presence of discontinuous transitions in internal state..."

Parallel to this line of research, but without invocation of dynamic systems theory, is what Adams (2003) has characterized as 'the informational turn in philosophy', that is, the application of communication theory formalism and concepts to "purposive behavior, learning, pattern recognition, and... the naturalization of mind and meaning". One of the first comprehensive attempts was that of Dretske (1981, 1988, 1992, 1993, 1994), whose work Adams describes as follows:

> "It is not uncommon to think that information is a commodity generated by things with minds. Let's say that a naturalized account puts matters the other way around, viz. it says that minds are things that come into being by purely natural causal means of exploiting the information in their environments. This is the approach of Dretske as he tried consciously to unite the cognitive sciences around the well-understood mathematical theory of communication..."

Dretske himself (1994) writes:

> "Communication theory can be interpreted as telling one something important about the conditions that are needed for the transmission of information as ordinarily understood, about what it takes for the transmission of semantic information. This has tempted people... to exploit [information theory] in semantic and cognitive studies, and thus in the philosophy of mind.

*Address correspondence to Rodrick Wallace, PISCS Inc., 549 W 123 St., Suite 16F, New York, NY, 10027. Telephone (212) 865-4766, email rdwall@ix.netcom.com. Affiliation is for identification only.

...Unless there is a statistically reliable channel of communication between [a source and a receiver]... no signal can carry semantic information... [thus] the channel over which the [semantic] signal arrives [must satisfy] the appropriate statistical constraints of communication theory."

Here we redirect attention from the informational content or meaning of individual symbols, i.e. the province of semantics which so concerned Dretske, back to the statistical properties of long, internally-structured paths of symbols emitted by an information source. We will then import a variety of tools from statistical physics to produce dynamically tunable punctuated or phase transition coupling between interacting cognitive modules in what we claim is a highly natural manner. As Dretske so clearly saw, this approach allows scientific inference on the necessary conditions for cognition, and, we will show, greatly illuminates dynamic neuronal workspace models of consciousness without raising the 18th Century ghosts of noisy, distorted mechanical clocks inherent to dynamic systems theory. It also opens the way for an extended dynamic workspace model which includes the effects of other interacting cognitive modules or embedding contexts that may, although acting at slower timescales, profoundly affect individual consciousness. This extension meets profound objections to brain-only models, for example those of Bennett and Hacker (2003), which we will consider in more detail below.

Before entering the formal thicket, it is important to highlight several points.

First, information theory is notorious for providing existence theorems whose representation, to use physics jargon, is arduous. For example, although the Shannon Coding Theorem implied the possibility of highly efficient coding schemes as early as 1949, it took more than forty years for practical 'turbo codes' to actually be constructed. The research program we implicitly propose here is unlikely to be any less difficult.

Second, we are invoking information theory variants of the fundamental limit theorems of probability. These are independent of exact mechanisms, but constrain the behavior of those mechanisms. For example, although not all processes involve long sums of independent stochastic variables, those that do, regardless of the individual variable distribution, collectively follow a Normal distribution as a consequence of the Central Limit Theorem. Similarly, the games of chance in a Las Vegas casino are all quite different, but nonetheless the success of strategies for playing them is strongly and systematically constrained by the Martingale Theorem, regardless of game details. We similarly propose that languages-on-networks and languages-that-interact, as a consequence of the limit theorems of information theory, will inherently be subject to regularities of tunable punctuation and generalized Onsager relations, regardless of detailed mechanisms, as important as the latter may be.

Just as parametric statistics are imposed, at least as a first approximation, on sometimes questionable experimental situations, relying on the robustness of the Central Limit Theorem to carry us through, we will pursue a similar heuristic approach here.

Finally, we invoke an obvious homology between information source uncertainty and thermodynamic free energy density as justification for importing renormalization and generalized Onsager relation formalism to the study of cognitive process near and away from 'critical points' in the coupling of cognitive submodules. The question of whether we are demonstrating the necessity of global phase transitions in information-transmission networks or merely building a suggestive analogy with thermodynamics is an empirical one beyond our present ability to answer, although for the microscopic case, Feynman (1996) has shown that the homology is an identity, which is no small matter and indeed suggests that phase transition behavior should be ubiquitous for certain classes of information systems.

Our work is likely analogous, in a certain sense, to Bohr's treatment of the atom, which attempted a simple substitution of quantized angular momentum into a basically classical theory. Although incomplete, that analysis contributed materially to the more comprehensive approaches of quantum mechanics, relativistic quantum mechanics, and quantum electrodynamics. In that spirit we hope that increasingly satisfactory models will follow from what we do here.

We begin with a description of cognitive process in terms of an information source, a kind of language constrained by the Shannon-McMillan or Asymptotic Equipartition Theorem, and its Rate Distortion or Joint Asymptotic Equipartition and other variants for interacting sources.

**Cognition as language**

Atlan and Cohen (1998) and Cohen (2000), following a long tradition in the study of immune cognition (e.g., Grossman, 1989; Tauber, 1998), argue that the essence of cognitive function involves comparison of a perceived signal with an internal, learned picture of the world, and then, upon that comparison, the choice of a response from a much larger repertoire of possible responses. Following the approach of Wallace (2000, 2002a), we make a 'weak', and hence very general, model of that process.

Cognitive pattern recognition-and-response, as we characterize it, proceeds by convoluting an incoming external sensory incoming signal with an internal ongoing activity – the learned picture of the world – and triggering an appropriate action based on a decision that the pattern of sensory activity requires a response. We will, fulfilling Atlan and Cohen's (1998) criterion of meaning-from-response, define a language's contextual meaning entirely in terms of system output, leaving out, for the moment, the question of how such a pattern recognition system is trained, a matter for Rate Distortion theory.

The abstract model will be illustrated by two neural network examples.

A pattern of sensory input is mixed in some unspecified but systematic manner with internal 'ongoing' activity to create a path of convoluted signal $x = (a_0, a_1, ..., a_n, ...)$. This path is fed into a highly nonlinear, but otherwise similarly unspecified, decision oscillator which generates an output $h(x)$ that is an element of one of two (presumably) disjoint sets $B_0$ and $B_1$ of possible system responses. We take

$$B_0 \equiv b_0, ..., b_k,$$

$$B_1 \equiv b_{k+1}, ..., b_m.$$

Thus we permit a graded response, supposing that if

$$h(x) \in B_0$$

the pattern is not recognized, and if

$$h(x) \in B_1$$

the pattern is recognized and some action $b_j, k+1 \leq j \leq m$ takes place.

We are interested in paths $x$ which trigger pattern recognition-and-response exactly once. That is, given a fixed initial state $a_0$, such that $h(a_0) \in B_0$, we examine all possible subsequent paths $x$ beginning with $a_0$ and leading exactly once to the event $h(x) \in B_1$. Thus $h(a_0, ..., a_j) \in B_0$ for all $j < m$, but $h(a_0, ..., a_m) \in B_1$.

For each positive integer $n$ let $N(n)$ be the number of paths of length $n$ which begin with some particular $a_0$ having $h(a_0) \in B_0$ and lead to the condition $h(x) \in B_1$. We shall call such paths 'meaningful' and assume $N(n)$ to be considerably less than the number of all possible paths of length $n$ – pattern recognition-and-response is comparatively rare. We further assume that the finite limit

$$H \equiv \lim_{n \to \infty} \frac{\log[N(n)]}{n}$$

both exists and is independent of the path $x$. We will – not surprisingly – call such a pattern recognition-and-response cognitive process *ergodic*. Not all such processes are likely to be ergodic, implying that $H$, if it exists, is path dependent, although extension to 'nearly' ergodic processes is straightforward.

Invoking Shannon, we may thus define an ergodic information source **X** associated with stochastic variates $X_j$ having joint and conditional probabilities $P(a_0, ..., a_n)$ and $P(a_n|a_0, ..., a_{n-1})$ such that appropriate joint and conditional Shannon uncertainties may be defined which satisfy the relations

$$H[\mathbf{X}] = \lim_{n \to \infty} \frac{\log[N(n)]}{n} =$$

$$\lim_{n \to \infty} H(X_n|X_0, ..., X_{n-1}) =$$

$$\lim_{n \to \infty} \frac{H(X_0, ..., X_n)}{n}.$$

(1)

The Shannon uncertainties $H(...)$ are defined in terms of cross-sectional sums of the form $-\sum_k P_k \log[P_k]$, where the $P_k$ constitute a probability distribution. See Ash (1990) or Cover and Thomas (1991) for details.

We say this information source is *dual* to the ergodic cognitive process.

Again, for non-ergodic sources, a limit $\lim_{n \to \infty} H$ may be defined for each path, but it will not necessarily given by the simple cross-sectional law-of-large numbers analogs above. For 'nearly' ergodic systems one might perhaps use something of the form

$$H(x + \delta x) \approx H(x) + \delta x dH/dx.$$

Different language-analogs will, of course, be defined by different divisions of the total universe of possible responses into different pairs of sets $B_0$ and $B_1$, or by requiring more than one response in $B_1$ along a path. However, like the use of different distortion measures in the Rate Distortion Theorem (e.g. Cover and Thomas, 1991), it seems obvious that the underlying dynamics will all be qualitatively similar.

Similar but not identical, and herein lies the first of two essential matters: dividing the full set of possible responses into sets $B_0$ and $B_1$ may itself require higher order cognitive decisions by another module or modules, suggesting the necessity of 'choice' within a more or less broad set of possible languages-of-thought. This would, in one way, reflect the need of the organism to shift gears according to the different challenges it faces, leading to a model for autocognitive disease when a normally excited state is recurrently (and incorrectly) identified as a member of the 'resting' set $B_0$.

A second possible source of structure, however, lies at the input rather than the output end of the model: i.e. suppose we classify paths instead of outputs. That is, we define equivalence classes in convolutional 'path space' according to whether a state $a_k^M$ can be connected by a path with some originating state $a_M$. That is, we, in turn, set each possible state to an $a_0$, and define other states as formally equivalent to it if they can be reached from that (now variable) $a_0 = a_M$ by a grammatical/syntactical path. That is, a state which can be reached by a legitimate path from $a_M$ is taken as equivalent to it. We can thus divide path space into (ordinarily) disjoint sets of equivalence classes. Each equivalence class defines its own language-of-thought: disjoint cognitive modules, possibly associated with an embedding equivalence class algebra.

While meaningful paths – creating an inherent grammar and syntax – are defined entirely in terms of system response, as Atlan and Cohen (1998) propose, a critical task is to make these (relatively) disjoint cognitive modules interact, and to examine the effects of that interaction on global properties. Punctuated phase transition effects will emerge in a natural manner.

Before proceeding, however, we give two explicit neural network applications.

First the simple stochastic neuron: A series of inputs $y_i^j, i = 1...m$ from $m$ nearby neurons at time $j$ is convoluted with 'weights' $w_i^j, i = 1...m$, using an inner product

$$a_j = \mathbf{y}^j \cdot \mathbf{w}^j = \sum_{i=1}^{m} y_i^j w_i^j$$

in the context of a 'transfer function' $f(\mathbf{y}^j \cdot \mathbf{w}^j)$ such that the probability of the neuron firing and having a discrete output $z^j = 1$ is $P(z^j = 1) = f(\mathbf{y}^j \cdot \mathbf{w}^j)$. Thus the probability that the neuron does not fire at time $j$ is $1 - f(\mathbf{y}^j \cdot \mathbf{w}^j)$.

In the terminology of this section the $m$ values $y_i^j$ constitute 'sensory activity' and the $m$ weights $w_i^j$ the 'ongoing activity' at time $j$, with $a_j = \mathbf{y}^j \cdot \mathbf{w}^j$ and $x = a_0, a_1, ... a_n, ...$

A little more work leads to a fairly standard neural network model in which the network is trained by appropriately varying the $\mathbf{w}$ through least squares or other error minimization feedback. This can be shown to, essentially, replicate rate distortion arguments (Cover and Thomas, 1991), as we can use the error definition to define a distortion function $d(y, \hat{y})$ which measures the difference between the training pattern $y$ and the network output $\hat{y}$ as a function of, for example, the inverse number of training cycles, $K$. As discussed in some detail elsewhere (Wallace, 2002), learning plateau behavior follows as a phase transition on the parameter $K$ in the mutual information $I(Y, \hat{Y})$.

Park et al. (2000) treat the stochastic neural network in terms of a space of related probability density functions $[p(\mathbf{x}, \mathbf{y}; \mathbf{w}) | \mathbf{w} \in \mathcal{R}^m]$, where $\mathbf{x}$ is the input, $\mathbf{y}$ the output and $\mathbf{w}$ the parameter vector. The goal of learning is to find an optimum $\mathbf{w}^*$ which maximizes the log likelihood function. They define a loss function of learning as

$$L(\mathbf{x}, \mathbf{y}; \mathbf{w}) \equiv -\log p(\mathbf{x}, \mathbf{y}; \mathbf{w}),$$

and one can take as a learning paradigm the gradient relation

$$\mathbf{w}_{t+1} = \mathbf{w}_t - \eta_t \partial L(\mathbf{x}, \mathbf{y}; \mathbf{w})/\partial \mathbf{w},$$

where $\eta_t$ is a learning rate.

Park et al. (2000) attack this optimization problem by recognizing that the space of $p(\mathbf{x}, \mathbf{y}; \mathbf{w})$ is Riemannian with a metric given by the Fisher information matrix

$$G(\mathbf{w}) = \int \int \partial \log p/\partial \mathbf{w} [\partial \log p/\partial \mathbf{w}]^T p(\mathbf{x}, \mathbf{y}; \mathbf{w}) d\mathbf{y} d\mathbf{x}$$

where $T$ is the transpose operation. A Fisher-efficient on-line estimator is then obtained by using the 'natural' gradient algorithm

$$\mathbf{w}_{t+1} = \mathbf{w}_t - \eta_t G^{-1} \partial L(\mathbf{x}, \mathbf{y}; \mathbf{w})/\partial \mathbf{w}.$$

Again, through the synergistic family of probability distributions $p(\mathbf{x}, \mathbf{y}; \mathbf{w})$, this can be viewed as a special case – a 'representation', to use physics jargon – of the general 'convolution argument' given above.

It seems likely that a rate distortion analysis of the interaction between training language and network response language will nonetheless show the ubiquity of learning plateaus, even in this rather elegant special case.

We will eventually parametize the information source uncertainty of the dual information source with respect to one or more variates, writing, e.g. $H[\mathbf{K}]$, where $\mathbf{K} \equiv (K_1, ..., K_s)$ represents a vector in a parameter space. Let the vector $\mathbf{K}$ follow some path in time, i.e. trace out a generalized line or surface $\mathbf{K}(t)$. We will, following the argument of Wallace (2002b), assume that the probabilities defining $H$, for the most part, closely track changes in $\mathbf{K}(t)$, so that along a particular 'piece' of a path in parameter space the information source remains as close to memoryless and ergodic as is needed for the mathematics to work. Between pieces, below, we will impose phase transition characterized by a renormalization symmetry, in the sense of Wilson (1971).

We will call such an information source 'adiabatically piecewise memoryless ergodic' (APME).

To anticipate the argument, iterating the analysis on paths of 'tuned' sets of renormalization parameters gives a second order punctuation in the rate at which primary interacting information sources representing cognitive submodules become linked to each other: the shifting workspace structure of consciousness.

### Interacting cognitive modules

We suppose that a two (relatively) distinct cognitive submodules can be represented by two distinct sequences of states, the paths $x \equiv x_0, x_1, ...$ and $y \equiv y_0, y_1, ....$ These paths are, however, both very highly structured and serially correlated and have dual information sources $\mathbf{X}$ and $\mathbf{Y}$. Since the modules, in reality, interact through some kind of endless back-and-forth mutual crosstalk, these sequences of states are not independent, but are jointly serially correlated. We can, then, define a path of sequential pairs as $z \equiv (x_0, y_0), (x_1, y_1), ....$ The essential content of the Joint Asymptotic Equipartition Theorem (JAEPT), a variant of the Shannon-McMillan Theorem, is that the set of joint paths $z$ can be partitioned into a relatively small set of high probability termed *jointly typical*, and a much larger set of vanishingly small probability. Further, according to the JAEPT, the *splitting criterion* between high and low probability sets of pairs is the mutual information

$$I(X, Y) = H(X) - H(X|Y) = H(X) + H(Y) - H(X, Y)$$

where $H(X), H(Y), H(X|Y)$ and $H(X, Y)$ are, respectively, the (cross-sectional) Shannon uncertainties of $X$ and $Y$, their conditional uncertainty, and their joint uncertainty. See Cover and Thomas (1991) for mathematical details. Similar approaches to neural process have been recently adopted by Dimitrov and Miller (2001).

Note that, using this asymptotic limit theorem approach, we need not model the exact form or dynamics of the crosstalk feedback, hence crushing algebraic complexities can be postponed until a later stage of the argument. They will, however, appear in due course with some vengeance.

The high probability pairs of paths are, in this formulation, all equiprobable, and if $N(n)$ is the number of jointly typical pairs of length $n$, then

$$I(X, Y) = \lim_{n \to \infty} \frac{\log[N(n)]}{n}.$$

Extending the earlier language-on-a-network models of Wallace and Wallace (1998, 1999), we suppose there is a coupling parameter $P$ representing the degree of linkage between the modules, and set $K = 1/P$, following the development of those earlier studies. Note that in a brain model this parameter represents the intensity of coupling between distant neural structures.

Then we have

$$I[K] = \lim_{n \to \infty} \frac{\log[N(K,n)]}{n}.$$

The essential 'homology' between information theory and statistical mechanics lies in the similarity of this expression with the infinite volume limit of the free energy density. If $Z(K)$ is the statistical mechanics partition function derived from the system's Hamiltonian, then the free energy density is determined by the relation

$$F[K] = \lim_{V \to \infty} \frac{\log[Z(K)]}{V}.$$

$F$ is the free energy density, $V$ the system volume and $K = 1/T$, where $T$ is the system temperature.

We and others argue at some length (Wallace and Wallace, 1998, 1999; Rojdestvensky and Cottam, 2000) that this is indeed a systematic mathematical homology which, we contend, permits importation of renormalization symmetry into information theory. Imposition of invariance under renormalization on the mutual information splitting criterion $I(X,Y)$ implies the existence of phase transitions analogous to learning plateaus or punctuated evolutionary equilibria. An extensive mathematical development will be presented in the next section.

The physiological details of mechanism, we speculate, will be particularly captured by the definitions of coupling parameter, renormalization symmetry, and, perhaps, the distribution of the renormalization across agency, a matter we treat below.

Here, however, these changes are perhaps better described as 'punctuated interpenetration' between interacting cognitive modules.

We reiterate that the details are highly dependent on the choice of renormalization symmetry (and its distribution), which is likely to reflect details of mechanism – the manner in which the dynamics of the forest are dependent on the detailed physiology of trees, albeit in a many-to-one manner. Renormalization properties are not likely to follow simple physical analogs, and may well be subject, in addition to complications of distribution, to the 'tuning' of universality class parameters that are characteristically fixed for simple physical systems. The algebra is straightforward if complicated, and given later.

### Representations of the general argument

**1. Language-on-a-network models.** Earlier papers of this series addressed the problem of how a language, in a large sense, spoken on a network structure responds as properties of the network change. The language might be speech, pattern recognition, or cognition. The network might be social, chemical, or neural. The properties of interest were the magnitude of 'strong' or 'weak' ties which, respectively, either disjointly partitioned the network or linked it across such partitioning. These would be analogous to local and mean-field couplings in physical systems.

We fix the magnitude of strong ties – to reiterate, those which disjointly partition the underlying network (presumably into cognitive submodules) – but vary the index of weak ties between components, which we call $P$, taking $K = 1/P$.

For interacting neural networks $P$ might simply be taken as proportional to the degree of crosstalk.

We assume the piecewise, adiabatically memoryless ergodic information source (Wallace, 2002b) depends on three parameters, two explicit and one implicit. The explicit are $K$ as above and an 'external field strength' analog $J$, which gives a 'direction' to the system. We will, in the limit, set $J = 0$.

The implicit parameter, which we call $r$, is an inherent generalized 'length' characteristic of the phenomenon, on which $J$ and $K$ are defined. That is, we can write $J$ and $K$ as functions of averages of the parameter $r$, which may be quite complex, having nothing at all to do with conventional ideas of space: For example $r$ may be defined by the degree of niche partitioning in ecosystems or separation in social structures.

For a given generalized language of interest with a well defined (piecewise adiabatically memoryless) ergodic source uncertainty $H$ we write

$$H[K, J, \mathbf{X}]$$

Imposition of invariance of $H$ under a renormalization transform in the implicit parameter $r$ leads to expectation of both a critical point in $K$, which we call $K_C$, reflecting a phase transition to or from collective behavior across the entire array, and of power laws for system behavior near $K_C$. Addition of other parameters to the system, e.g. some $V$, results in a 'critical line' or surface $K_C(V)$.

Let $\kappa \equiv (K_C - K)/K_C$ and take $\chi$ as the 'correlation length' defining the average domain in $r$-space for which the information source is primarily dominated by 'strong' ties. We begin by averaging across $r$-space in terms of 'clumps' of length $R$. Then, taking Wilson's (1971) analysis as a starting point, we choose the renormalization relations as

$$H[K_R, J_R, \mathbf{X}] = f(R) H[K, J, \mathbf{X}]$$

$$\chi(K_R, J_R) = \frac{\chi(K, J)}{R},$$

(2)

with $f(1) = 1$ and $J_1 = J, K_1 = K$. The first of these equations significantly extends Wilson's treatment. It states that 'processing capacity,' as indexed by the source uncertainty of the system, representing the 'richness' of the generalized language, grows monotonically as $f(R)$, which must itself be a dimensionless function in $R$, since both $H[K_R, J_R]$ and $H[K, J]$ are themselves dimensionless. Most simply, this would require that we replace $R$ by $R/R_0$, where $R_0$ is the 'characteristic length' for the system over which renormalization procedures are reasonable, then set $R_0 \equiv 1$, i.e. measure length in units of $R_0$. Wilson's original analysis focused on free energy density. Under 'clumping', densities must remain the same, so that if $F[K_R, J_R]$ is the free energy of the clumped system, and $F[K, J]$ is the free energy density before clumping, then Wilson's equation (4) is $F[K, J] = R^{-3} F[K_R, J_R]$, i.e.

$$F[K_R, J_R] = R^3 F[K, J].$$

Remarkably, the renormalization equations are solvable for a broad class of functions $f(R)$, or more precisely, $f(R/R_0), R_0 \equiv 1$.

The second relation just states that the correlation length simply scales as $R$.

Other, very subtle, symmetry relations – not necessarily based on the elementary physical analog we use here – may well be possible. For example McCauley, (1993, p.168) describes the highly counterintuitive renormalization relations needed to understand phase transition in simple 'chaotic' systems. This is important, since we suspect that biological or social systems may alter their renormalization properties – equivalent to tuning their phase transition dynamics – in response to external signals. We will make much of this possibility, termed 'universality class tuning', below.

To begin, following Wilson, we take $f(R) = R^d$ for some real number $d > 0$, and restrict $K$ to near the 'critical value' $K_C$. If $J \to 0$, a simple series expansion and some clever algebra (Wilson, 1971; Binney et al., 1986) gives

$$H = H_0 \kappa^\alpha$$

$$\chi = \frac{\chi_0}{\kappa^s}$$

(3)

where $\alpha, s$ are positive constants. We provide more biologically relevant examples below.

Further from the critical point matters are more complicated, appearing to involve Generalized Onsager Relations and a kind of thermodynamics associated with a Legendre transform of $H$, i.e. $S \equiv H - K dH/dK$ (Wallace, 2002a). Although this extension is quite important to describing behaviors away from criticality, the full mathematical detail is cumbersome and the reader is referred to the references. A brief discussion will be given below.

An essential insight is that *regardless of the particular renormalization properties, sudden critical point transition is possible in the opposite direction for this model*. That is, we go from a number of independent, isolated and fragmented systems operating individually and more or less at random, into a single large, interlocked, coherent structure, once the parameter $K$, the inverse strength of weak ties, falls below threshold, or, conversely, once the strength of weak ties parameter $P = 1/K$ becomes large enough.

Thus, increasing nondisjunctive weak ties between them can bind several different cognitive 'language' functions into a single, embedding hierarchical metalanguage which contains each as a linked subdialect, and do so in an inherently punctuated manner. This could be a dynamic process, creating a shifting, ever-changing, pattern of linked cognitive submodules, according to the challenges or opportunities faced by the organism.

To reiterate somewhat, this heuristic insight can be made more exact using a rate distortion argument (or, more generally, using the Joint Asymptotic Equipartition Theorem) as follows (Wallace, 2002a, b):

Suppose that two ergodic information sources **Y** and **B** begin to interact, to 'talk' to each other, i.e. to influence each other in some way so that it is possible, for example, to look at the output of **B** – strings $b$ – and infer something about the behavior of **Y** from it – strings $y$. We suppose it possible to define a retranslation from the B-language into the Y-language through a deterministic code book, and call $\hat{\mathbf{Y}}$ the translated information source, as mirrored by **B**.

Define some distortion measure comparing paths $y$ to paths $\hat{y}$, $d(y, \hat{y})$ (Cover and Thomas, 1991). We invoke the Rate Distortion Theorem's mutual information $I(Y, \hat{Y})$, which is the splitting criterion between high and low probability pairs of paths. Impose, now, a parametization by an inverse coupling strength $K$, and a renormalization symmetry representing the global structure of the system coupling. This may be much different from the renormalization behavior of the individual components. If $K < K_C$, where $K_C$ is a critical point (or surface), the two information sources will be closely coupled enough to be characterized as condensed.

In the absence of a distortion measure, we can invoke the Joint Asymptotic Equipartition Theorem to obtain a similar result.

We suggest in particular that detailed coupling mechanisms will be sharply constrained through regularities of grammar and syntax imposed by limit theorems associated with phase transition.

Wallace and Wallace (1998, 1999) and Wallace (2002) use this approach to address certain evolutionary processes in a relatively unified fashion. These papers, and those of Wallace and Fullilove (1999) and Wallace (2002a), further describe how biological or social systems might respond to gradients in information source uncertainty and related quantities when the system is away from phase transition. Language-on-network systems, as opposed to physical systems, appear to diffuse away from concentrations of an 'instability' construct which is related to a Legendre transform of information source uncertainty, in much the same way entropy is the Legendre transform of free energy density in a physical system.

Simple thermodynamics addresses physical systems held at or near equilibrium conditions. Treatment of nonequilibrium, for example highly dynamic, systems requires significant extension of thermodynamic theory. The most direct approach has been the first-order phenomenological theory of Onsager, which involves relating first order rate changes in system parameters $K_j$ to gradients in physical entropy $S$, involving 'Onsager relation' equations of the form

$$\sum_k R_{k,j} dK_j/dt = \partial S/\partial K_j,$$

where the $R_{k,j}$ are characteristic constants of a particular system and $S$ is defined to be the Legendre transform free energy density $F$;

$$S \equiv F - \sum_j \partial F/\partial K_j.$$

The entropy-analog for an information system is, then, the dimensionless quantity

$$S \equiv H - \sum_j K_j \partial H/\partial K_j,$$

or a similar equation in the mutual information $I$.

Note that in this treatment $I$ or $H$ play the role of free energy, not entropy, and that their Legendre transform plays the role of physical entropy. This is a key matter.

For information systems, a parametized 'instability', $Q[K] \equiv S - H$, is defined from the principal splitting criterion by the relations

$$Q[K] = -KdH[K]/dK$$

$$Q[K] = -KdI[K]/dK$$

(4)

where $H[K]$ and $I[K]$ are, respectively, information source uncertainty or mutual information in the Asymptotic Equipartition, Rate Distortion, or Joint Asymptotic Equipartition Theorems.

Extension of thermodynamic theory to information systems involves a first order system phenomenological equations analogous to the Onsager relations, but possibly having very complicated behavior in the $R_{j,k}$, in particular not necessarily producing simple diffusion toward peaks in $S$. For example, as discussed, there is evidence that social network structures are affected by diffusion *away* from concentrations in the S-analog. Thus the phenomenological relations affecting the dynamics of information networks, which are inherently open systems, may not be governed simply by mechanistic diffusion toward 'peaks in entropy', but may, in first order, display more complicated behavior.

**2. 'Biological' phase transitions.** Now the mathematical detail concealed by the invocation of the asymptotic limit theorems emerges with a vengeance. Equation (2) states that the information source and the correlation length, the degree of coherence on the underlying network, scale under renormalization clustering in chunks of size $R$ as

$$H[K_R, J_R]/f(R) = H[J, K]$$

$$\chi[K_R, J_R]R = \chi(K, J),$$

with $f(1) = 1, K_1 = K, J_1 = J$, where we have slightly rearranged terms.

Differentiating these two equations with respect to $R$, so that the right hand sides are zero, and solving for $dK_R/dR$ and $dJ_R/dR$ gives, after some consolidation, expressions of the form

$$dK_R/dR = u_1 d\log(f)/dR + u_2/R$$

$$dJ_R/dR = v_1 J_R d\log(f)/dR + \frac{v_2}{R} J_R.$$

(5)

The $u_i, v_i, i = 1, 2$ are functions of $K_R, J_R$, but not explicitly of $R$ itself.

We expand these equations about the critical value $K_R = K_C$ and about $J_R = 0$, obtaining

$$dK_R/dR = (K_R - K_C)y d\log(f)/dR + (K_R - K_C)z/R$$

$$dJ_R/dR = wJ_R d\log(f)/dR + xJ_R/R.$$

(6)

The terms $y = du_1/dK_R|_{K_R=K_C}, z = du_2/dK_R|_{K_R=K_C}, w = v_1(K_C, 0), x = v_2(K_C, 0)$ are constants.

Solving the first of these equations gives

$$K_R = K_C + (K - K_C)R^z f(R)^y,$$

(7)

again remembering that $K_1 = K, J_1 = J, f(1) = 1$.

Wilson's essential trick is to iterate on this relation, which is supposed to converge rapidly (Binney, 1986), assuming that for $K_R$ near $K_C$, we have

$$K_C/2 \approx K_C + (K - K_C)R^z f(R)^y.$$

(8)

We iterate in two steps, first solving this for $f(R)$ in terms of known values, and then solving for $R$, finding a value $R_C$ that we then substitute into the first of equations (2) to obtain an expression for $H[K, 0]$ in terms of known functions and parameter values.

The first step gives the general result

$$f(R_C) \approx \frac{[KC/(KC-K)]^{1/y}}{2^{1/y} R_C^{z/y}}. \tag{9}$$

Solving this for $R_C$ and substituting into the first of equation (2) gives, as a first iteration of a far more general procedure (e.g. Shirkov and Kovalev, 2001)

$$H[K,0] \approx \frac{H[K_C/2,0]}{f(R_C)} = \frac{H_0}{f(R_C)}$$

$$\chi(K,0) \approx \chi(K_C/2,0)R_C = \chi_0 R_C \tag{10}$$

which are the essential relationships.

Note that a power law of the form $f(R) = R^m, m = 3$, which is the direct physical analog, may not be biologically reasonable, since it says that 'language richness' can grow very rapidly as a function of increased network size. Such rapid growth is simply not observed.

If we take the biologically realistic example of non-integral 'fractal' exponential growth,

$$f(R) = R^\delta, \tag{11}$$

where $\delta > 0$ is a real number which may be quite small, we can solve equation (8) for $R_C$, obtaining

$$R_C = \frac{[KC/(KC-K)]^{[1/(\delta y+z)]}}{2^{1/(\delta y+z)}} \tag{12}$$

for $K$ near $K_C$. Note that, for a given value of $y$, we might want to characterize the relation $\alpha \equiv \delta y + z =$ constant as a "tunable universality class relation" in the sense of Albert and Barabasi (2002).

Substituting this value for $R_C$ back into equation (9) gives a somewhat more complex expression for $H$ than equation (2), having three parameters, i.e. $\delta, y, z$.

A more biologically interesting choice for $f(R)$ is a logarithmic curve that 'tops out', for example

$$f(R) = m\log(R) + 1. \tag{13}$$

Again $f(1) = 1$.
Using Mathematica 4.2 to solve equation (8) for $R_C$ gives

$$R_C = [\frac{Q}{LambertW[Q\exp(z/my)]}]^{y/z}, \tag{14}$$

where

$$Q \equiv [(z/my)2^{-1/y}[KCKC-K]]^{1/y}.$$

The transcendental function LambertW(x) is defined by the relation

$$LambertW(x)\exp(LambertW(x)) = x.$$

It arises in the theory of random networks and in renormalization strategies for quantum field theories.

An asymptotic relation for $f(R)$ would be of particular biological interest, implying that 'language richness' increases to a limiting value with population growth. Such a pattern is broadly consistent with calculations of the degree of allelic heterozygosity as a function of population size under a balance between genetic drift and neutral mutation (Hartl and Clark, 1997; Ridley, 1996). Taking

$$f(R) = \exp[m(R-1)/R] \tag{15}$$

gives a system which begins at 1 when R=1, and approaches the asymptotic limit $\exp(m)$ as $R \to \infty$. Mathematica 4.2 finds

$$R_C = \frac{my/z}{LambertW[S]} \tag{16}$$

where

$$S \equiv (my/z)\exp(my/z)[2^{1/y}[KC/(KC-K)]^{-1/y}]^{y/z}.$$

These developments indicate the possibility of taking the theory significantly beyond arguments by abduction from simple physical models, although the notorious difficulty of implementing information theory existence arguments will undoubtedly persist.

**3. Universality class distribution.** Physical systems undergoing phase transition usually have relatively pure renormalization properties, with quite different systems clumped into the same 'universality class', having fixed exponents at transition (e.g. Binney, 1986). Biological and social phenomena may be far more complicated:

If we suppose the system of interest to be a mix of subgroups with different values of some significant renormalization parameter $m$ in the expression for $f(R, m)$, according to a distribution $\rho(m)$, then we expect the first expression in equation (1) to generalize as

$$H[K_R, J_R] = <f(R,m)> H[K, J]$$
$$\equiv H[K, J] \int f(R,m)\rho(m)dm.$$
(17)

If $f(R) = 1 + m \log(R)$ then, given any distribution for $m$, we simply obtain

$$<f(R)> = 1+ <m> \log(R)$$
(18)

where $<m>$ is simply the mean of $m$ over that distribution.

Other forms of $f(R)$ having more complicated dependencies on the distributed parameter or parameters, like the power law $R^\delta$, do not produce such a simple result. Taking $\rho(\delta)$ as a normal distribution, for example, gives

$$<R^\delta> = R^{<\delta>} \exp[(1/2)(\log(R^\sigma))^2],$$
(19)

where $\sigma^2$ is the distribution variance. The renormalization properties of this function can be determined from equation (8), and is left to the reader as an exercise, best done in Mathematica 4.2 or above.

Thus the information dynamic phase transition properties of mixed systems will not in general be simply related to those of a single subcomponent, a matter of possible empirical importance: If sets of relevant parameters defining renormalization universality classes are indeed distributed, experiments observing pure phase changes may be very difficult. Tuning among different possible renormalization strategies in response to external pressures would result in even greater ambiguity in recognizing and classifying information dynamic phase transitions.

We believe that important aspects of mechanism may be reflected in the combination of renormalization properties and the details of their distribution across subsystems.

In sum, real biological, social, or interacting biopsychosocial systems are likely to have very rich patterns of phase transition which may not display the simplistic, indeed, literally elemental, purity familiar to physicists. Overall mechanisms will, we believe, still remain significantly constrained by our theory, in the general sense of probability limit theorems.

**4. Universality class tuning: the fluctuating dynamic threshold** Next we iterate the general argument onto the process of phase transition itself, producing our model of consciousness as a tunable neural workspace subject to inherent punctuated detection of external events.

An essential character of physical systems subject to phase transition is that they belong to particular 'universality classes'. This means that the exponents of power laws describing behavior at phase transition will be the same for large groups of markedly different systems, with 'natural' aggregations representing fundamental class properties (e.g. Binney et al., 1986).

It is our contention that biological or social systems undergoing phase transition analogs need not be constrained to such classes, and that 'universality class tuning', meaning the strategic alteration of parameters characterizing the renormalization properties of punctuation, might well be possible. Here we focus on the tuning of parameters within a single, given, renormalization relation. Clearly, however, wholesale shifts of renormalization properties must ultimately be considered as well.

Universality class tuning has been observed in models of 'real world' networks. As Albert and Barabasi (2002) put it,

> "The inseparability of the topology and dynamics of evolving networks is shown by the fact that [the exponents defining universality class] are related by [a] scaling relation..., underlying the fact that a network's assembly uniquely determines its topology. However, in no case are these exponents unique. They can be tuned continuously..."

We suppose that a structured external environment, which we take itself to be an appropriately regular information source **Y** 'engages' a modifiable cognitive system. The environment begins to write an image of itself on the cognitive system in a distorted manner permitting definition of a mutual information $I[K]$ splitting criterion according to the Rate Distortion or Joint Asymptotic Equipartition Theorems. $K$ is an inverse coupling parameter between system and environment (Wallace, 2002a, b). According to our development, at

punctuation – near some critical point $K_C$ – the systems begin to interact very strongly indeed, and we may write, near $K_C$, taking as the starting point the simple physical model of equation (2),

$$I[K] \approx I_0 [\frac{K_C - K}{K_C}]^\alpha.$$

For a physical system $\alpha$ is fixed, determined by the underlying 'universality class'. Here we will allow $\alpha$ to vary, and, in the section below, to itself respond explicitly to signals.

Normalizing $K_C$ and $I_0$ to 1, we obtain,

$$I[K] \approx (1-K)^\alpha.$$

(20)

The horizontal line $I[K] = 1$ corresponds to $\alpha = 0$, while $\alpha = 1$ gives a declining straight line with unit slope which passes through 0 at $K = 1$. Consideration shows there are progressively sharper transitions between the necessary zero value at $K = 1$ and the values defined by this relation for $0 < K, \alpha < 1$. The rapidly rising slope of transition with declining $\alpha$ is, we assert, of considerable significance.

The instability associated with the splitting criterion $I[K]$ is defined by

$$Q[K] \equiv -KdI[K]/dK = \alpha K(1-K)^{\alpha-1},$$

(21)

and is singular at $K = K_C = 1$ for $0 < \alpha < 1$. Following earlier work (Wallace and Wallace, 1998, 1999; Wallace and Fullilove, 1999; Wallace, 2002a), we interpret this to mean that values of $0 < \alpha \ll 1$ are highly unlikely for real systems, since $Q[K]$, in this model, represents a kind of barrier for information systems, in particular neural networks.

On the other hand, smaller values of $\alpha$ mean that the system is far more efficient at responding to the adaptive demands imposed by the embedding structured environment, since the mutual information which tracks the matching of internal response to external demands, $I[K]$, rises more and more quickly toward the maximum for smaller and smaller $\alpha$ as the inverse coupling parameter $K$ declines below $K_C = 1$. That is, *systems able to attain smaller $\alpha$ are more responsive to external signals than those characterized by larger values*, in this model, but smaller values will be hard to reach, and can probably be done so only at some considerable physiological or opportunity cost: focused conscious action takes resources, of one form or another.

The more biologically realistic renormalization strategies given above produce sets of several parameters defining the universality class, whose tuning gives behavior much like that of $\alpha$ in this simple example.

We can formally iterate the phase transition argument on this calculation to obtain our version of tunable consciousness, focusing on paths of universality class parameters.

Suppose the renormalization properties of a language-on-a network system at some 'time' $k$ are characterized by a set of parameters $A_k \equiv \alpha_1^k, ..., \alpha_m^k$. Fixed parameter values define a particular universality class for the renormalization. We suppose that, over a sequence of 'times', the universality class properties can be characterized by a path $x_n = A_0, A_1, ..., A_{n-1}$ having significant serial correlations which, in fact, permit definition of an adiabatically piecewise memoryless ergodic information source associated with the paths $x_n$. We call that source **X**.

We further suppose, in the manner of Wallace (2002a, b), that the set of external signals is also highly structured and forms another information source **Y** which interacts not only with the system of interest globally, but specifically with its universality class properties as characterized by **X**. **Y** is necessarily associated with a set of paths $y_n$.

We pair the two sets of paths into a joint path, $z_n \equiv (x_n, y_y)$ and invoke an inverse coupling parameter, $K$, between the information sources and their paths. This leads, by the arguments above, to phase transition punctuation of $I[K]$, the mutual information between **X** and **Y**, under either the Joint Asymptotic Equipartition Theorem or under limitation by a distortion measure, through the Rate Distortion Theorem (Cover and Thomas, 1991). Again, see Wallace (2002a, b) for more details of the argument. The essential point is that $I[K]$ is a splitting criterion under these theorems, and thus partakes of the homology with free energy density which we have invoked above.

Activation of universality class tuning, our version of attentional focusing, then becomes itself a punctuated event in response to increasing linkage between the organism and an external structured signal.

We note in passing, but without further calculation, that another path to the fluctuating dynamic threshold might be through a second order iteration similar to that just above, but focused on the parameters defining the universality class distributions of section 3.

### Extending the model

The Rate Distortion and Joint Asymptotic Equipartition Theorems are generalizations of the Shannon-McMillan Theorem which examine the interaction of two information sources, with and without the constraint of a fixed average distortion. We conduct one more iteration, and require a generalization of the SMT in terms of the splitting criterion for triplets as opposed to single or double stranded patterns. The tool for this is at the core of what is termed *network information theory* [Cover and Thomas, 1991, Theorem 14.2.3]. Suppose we have three (piecewise adiabatically memoryless) ergodic information sources, $Y_1, Y_2$ and $Y_3$. We assume $Y_3$ constitutes a critical embedding context for $Y_1$ and $Y_2$ so that, given three sequences of length $n$, the probability of a particular triplet of sequences is determined by *conditional probabilities with respect to $Y_3$*:

$$P(Y_1 = y_1, Y_2 = y_2, Y_3 = y_3) =$$

$$\Pi_{i=1}^n p(y_{1i}|y_{3i})p(y_{2i}|y_{3i})p(y_{3i}).$$

(22)

That is, $Y_1$ and $Y_2$ are, in some measure, driven by their interaction with $Y_3$

Then, in analogy with previous analyses, triplets of sequences can be divided by a splitting criterion into two sets, having high and low probabilities respectively. For large $n$ the number of triplet sequences in the high probability set will be determined by the relation [Cover and Thomas, 1992, p. 387]

$$N(n) \propto \exp[nI(Y_1; Y_2|Y_3)],$$

(23)

where splitting criterion is given by

$$I(Y_1; Y_2|Y_3) \equiv$$

$$H(Y_3) + H(Y_1|Y_3) + H(Y_2|Y_3) - H(Y_1, Y_2, Y_3)$$

We can then examine mixed cognitive/adaptive phase transitions analogous to learning plateaus (Wallace, 2002b) in the splitting criterion $I(Y_1, Y_2|Y_3)$, which characterizes the synergistic interaction between $Y_3$, taken as an embedding context, and the cognitive processes characterized by $Y_1$ and $Y_2$. These transitions delineate the various stages of the chronic infection, which are embodied in the slowly varying 'piecewise adiabatically memoryless ergodic' phase between transitions. Again, our results are exactly analogous to the Eldredge/Gould model of evolutionary punctuated equilibrium.

We can, if necessary, extend this model to any number of interacting information sources, $Y_1, Y_2, ..., Y_s$ conditional on an external context $Z$ in terms of a splitting criterion defined by

$$I(Y_1; ...; Y_s|Z) = H(Z) + \sum_{j=1}^{s} H(Y_j|Z) - H(Y_1, ..., Y_s, Z),$$

(24)

where the conditional Shannon uncertainties $H(Y_j|Z)$ are determined by the appropriate direct and conditional probabilities.

Note that this simple-seeming extension opens a Pandora's box in the study of 'mind-body interaction' and the impacts of culture on individual cognition. We now have a tool for examining the interpenetration of a broad range of cognitive physiological, psychological, and social submodules – not just neural substructures – with each other and with embedding contextual cultural language so characteristic of human hypersociality, all within the further context of structured psychosocial stress. Wallace (2003) analyzes the implications for understanding comorbid mind/body dysfunction, and provides a laundry list of physiological, psychological, and social cognitive modules associated with health and disease.

Bennett and Hacker (2003) define the 'mereological fallacy' in neuroscience as the assignment, to parts of an animal, of those characteristics which are properties of the whole. Humans, through both their embedding in cognitive social networks, and their secondary epigenetic inheritance system of culture, are even more than 'simply' individual animals. Equation (24) implies the possibility of extending the global neuronal workspace model of consciousness to include both internal cognitive physiological systems and embedding cognitive and other structures, providing a natural approach to evading that fallacy.

Equation (24) is itself subject to significant generalization. The single information source $Z$ is seen here as invariant, not affected by, but affecting, cross talk with the information sources for which it serves as the driving context. Suppose there is an interacting system of contexts, acting more slowly than the global neuronal workspace, but communicating within itself. It should be possible, at first order, to divide the full system into two sections, one 'fast', containing the $Y_j$, and the other 'slow', containing the series of information sources $Z_k$. The fast system instantiates the conscious neuronal workspace, including crosstalk, while the slow system constitutes an embedding context for the fast, but one which engages in its own pattern of crosstalk. Then the extended splitting criterion, which we write as

$$I(Y_1, ..., Y_j|Z_1, ..., Z_k),$$

becomes something far more complicated than equation (24). This must be expressed in terms of sums of appropriate Shannon uncertainties, a complex task which will be individually contingent on the particular forms of context and their interrelations.

**Discussion and conclusions**

We have constructed a punctuated information dynamic global neuronal workspace model which incorporates a second-order and similarly punctuated universality class tuning linked to detection of structured external signals. Tuning the punctuated activation of attention to those signals permits more rapid and appropriate response, but at increased physiological or other opportunity cost: unconscious processing is clearly more efficient, if the organism can get away with it. On the other hand, if the organism can't get away with it, death is more likely, suggesting a strong evolutionary imperative for a dynamic global neural workspace.

Linkage across individual dynamic workspaces – i.e. human hypersociality in the context of an embedding epigenetic cultural inheritance system – would be even more adaptationally efficient. Indeed, the last equation and its proposed extension suggest the possibility of very strong linkage of individual consciousness and physiology to embedding sociocultural network phenomena, ultimately producing an extended model of consciousness which does not fall victim to the mereological fallacy.

In just this regard Nisbett et al. (2001) review an extensive literature on empirical studies of basic cognitive differences between individuals raised in what they call 'East Asian' and 'Western' cultural heritages, which they characterize, respectively, as 'holistic' and 'analytic'. They find:

1. Social organization directs attention to some aspects of the perceptual field at the expense of others.

2. What is attended to influences metaphysics.

3. Metaphysics guides tacit epistemology, that is, beliefs about the nature of the world and causality.

4. Epistemology dictates the development and application of some cognitive processes at the expense of others.

5. Social organization can directly affect the plausibility of metaphysical assumptions, such as whether causality should be regarded as residing in the field vs. in the object.

6. Social organization and social practice can directly influence the development and use of cognitive processes such as dialectical vs. logical ones.

Nisbett et al. (2001) conclude that tools of thought embody a culture's intellectual history, that tools have theories built into them, and that users accept these theories, albeit unknowingly, when they use these tools.

Individual consciousness – currently defined in terms of the global neuronal workspace – appears to be profoundly affected by cultural, and perhaps developmental, context, and, we aver, by patterns of embedding psychosocial stress, all matters subject to a direct empirical study which may lead to an extension of the concept particularly useful in understanding certain forms of psychopathology.

From even limited theoretical perspectives, current dynamic systems models of neural networks, or their computer simulations, simply do not reflect the imperatives of Adams' (2003) informational turn in philosophy. On the other hand, dynamic systems models based on differential equations, or their difference equation realizations on computers, have a history of intense and continuous intellectual development going back to Isaac Newton. Hence very little new mathematics needs to be done, and one can look up most of the needed results in the textbooks, which are quite sophisticated by now. Rigorous probability theory is perhaps a hundred years old, and its information theory subset has seen barely a half century. Consequently the mathematics can't always be looked up, and sometimes must even be created de novo, at no small difficulty. One is reminded, not originally, of a drunk looking for his lost car keys under a street lamp 'because the light is better here'.

Virtually all self-proclaimed applications of information theory to the dynamic neural workspace currently in the neuroscience literature have strayed far indeed from the draconian structural discipline imposed by the asymptotic limit theorems of the subject: information measures are of no fundamental interest in and of themselves, serving only as grist for the mills of splitting criteria between high and low probability sets of dynamic paths. This is the central mechanism whose extension, using a homology with free energy density, permits exploration of punctuated neural dynamics in a manner consistent with the program described by Adams (2003).

According to the mathematical ecologist E.C. Pielou (1976, p.106), the legitimate purpose of mathematical models is to raise questions for empirical study, not to answer them, or, as one wag put it, "all models are wrong, but some models are useful". The natural emergence of tunable punctuated dynamics in our treatment, albeit at the expense of elaborate renormalization calculations at transition, and generalized Onsager relations away from it, suggests the possible utility of the theory in future empirical studies of consciousness: the car keys really may have been lost in the dark parking lot down the street, but here is a new flashlight.

We have outlined an empirically-testable approach to modeling consciousness which returns with a resounding thump to the classic asymptotic limit theorems of communication theory, and suggests further the necessity of incorporating the effects of embedding structures of psychosocial stress and culture. The theory suffers from a painful grandiosity, claiming to incorporate matters of cognition, consciousness, social system, and culture into a single all-encompassing model. To quote from a recent review of Bennett and Hacker's new book, (Patterson, 2003), however, contemporary neuroscience itself may suffer a more pernicious and deadly form of that disorder for which our approach is, in fact, the antidote:

> "[Bennett and Hacker] argue that for some neuroscientists, the brain does all manner of things: it believes (Crick); interprets (Edelman); knows (Blakemore); poses questions to itself (Young); makes decisions (Damasio); contains symbols (Gregory) and represents information (Marr). Implicit in these assertions is a philosophical mistake, insofar as it unreasonably inflates the conception of the 'brain' by assigning to it powers and activities that are normally reserved for sentient beings... these claims are not false; rather they are devoid of sense."

This is but one example of a swelling critical chorus which will grow markedly in virulence and influence. Our development, or some related version, leads toward explicit incorporation of the full 'sentient being' into observational studies of consciousness. For humans, whose hypersociality is both glory and bane, this particularly involves understanding the effects of embedding social and cultural context on immediate conscious experience.

The bottom line would seem to be the urgent necessity of extending the perspective of Nisbett et al. (2001) to brain imaging and other empirical studies of consciousness, and extending the global neuronal workspace model accordingly, a matter which our development here suggests is indeed possible, if not straightforward.